\documentclass[pdf]{iucr}              

\RequirePackage{graphicx}
\RequirePackage{amsmath,amsfonts,amsthm,bm,url}

     \journalcode{S}              

\begin{document}                  



\title{Tomographic Reconstruction using Tilted Laue Analyser-Based X-ray Phase-Contrast Imaging}


\cauthor[a]{M.C.}{Chalmers}{mcc1212@protonmail.com, JSR DOI: \url{https://doi.org/10.1107/S1600577520013995}}{address if different from \aff}
\author[b]{M.J.}{Kitchen}
\author[c]{K.}{Uesugi}
\author[d,e]{G.}{Falzon}
\author[e,f]{P.}{Quin}
\author[a,b,e]{K.M.}{Pavlov}

\aff[a]{University of Canterbury, 20 Kirkwood Avenue, Upper Riccarton, Christchurch 8041, New Zealand}
\aff[b]{Monash University, Wellington Rd, Clayton VIC 3800, Australia}
\aff[c]{Japan Synchrotron Radiation Research Institute, 1-1-1, Kouto, Sayo-cho, Sayo-gun, Hyogo 679-5198, Japan}
\aff[d]{College of Science and Engineering, Flinders University, Adelaide, South Australia, Australia, 5001}
\aff[e]{University of New England, Armidale NSW 2351, Australia}
\aff[f]{University of Tasmania, Churchill Ave, Hobart TAS 7005, Australia}








\numberwithin{equation}{section}
\maketitle                        

\begin{synopsis}
We applied an inclined geometry method to a Laue geometry setup for X-ray ABPCI through rotation of the detector and object about the optical axis allowing this traditionally 1-D phase sensitive phase contrast method to possess 2-D phase sensitivity. We acquired tomographic datasets over $360^\circ$ of a multi-material phantom to reconstruct the real and imaginary parts of the refractive index of the phantom.
\end{synopsis}

\begin{abstract}
Analyser-Based Phase-Contrast imaging (ABPCI) is a highly sensitive phase-contrast imaging method that produces high contrast images of weakly absorbing materials. However, it is only sensitive to phase gradient components lying in the diffraction plane of the analyser crystal (i.e in one dimension; 1-D). In order to accurately account for and measure phase effects produced by the wavefield-sample interaction, ABPCI and other 1-D phase sensitive methods must achieve 2-D phase gradient sensitivity. We applied an inclined geometry method to a Laue geometry setup for X-ray ABPCI through rotation of the detector and object about the optical axis. This allowed this traditionally 1-D phase sensitive phase contrast method to possess 2-D phase gradient sensitivity. We acquired tomographic datasets over $360^\circ$ of a multi-material phantom with the detector and sample tilted by $8^\circ$. The real and imaginary parts of the refractive index were reconstructed for the phantom.
\end{abstract}


\section{Introduction}

Phase-contrast x-ray imaging provides superior contrast for materials of low atomic number, including soft tissues, compared to traditional attenuation-based radiography, especially in the high energy regimes \cite{russo2017handbook}. This has the potential to enable greater image quality with less radiation dose delivered to the patient in a clinical setting \cite{keyrilainen2011,kitchen2017ct}. Analyser-Based Phase-Contrast imaging (ABPCI), also referred to as Diffraction Enhanced imaging, is a phase-contrast imaging technique that utilizes an analyser crystal to render phase gradients visible \cite{goetz1979,forster1980double,somenkov1991refraction,davis1995,bushuev1996dynamicalx,gureyev1997regimes,bushuev1997wave,chapman1997,bushuev1998wave,bravin2003exploiting,menk2005diffraction,coan2005phase,brankov2006computed,rigon2007generalized,zhou2014analyzer}. ABPCI is highly sensitive to components of phase gradients lying in the plane of diffraction of the analyser crystal, meaning it has 1-D phase sensitivity \cite{andre2001dynamical,wilkins2014}. The analyser crystal is mainly sensitive to the first derivative of the phase shift caused by the sample, which means it can pick up small discrepancies in the wavefield propagated through a sample. This 1-D phase sensitivity is also typical in other phase contrast methods such as grating interferometry \cite{david2002,momose2003}. For grating interferometry, \citeasnoun{ruthihauser2011} developed a method in a computed tomography setup to overcome the problem of 1-D sensitivity by utilizing an inclined geometry for the two 1-D gratings through rotation of $45^\circ$ about the optical axis to reconstruct a 2-D phase gradient. Taking a tomographic projection and its respective $180^\circ$ projection, then flipping the second projection enables orthogonal components of the phase gradient to be reconstructed. These can be combined and integrated to retrieve the phase map.
\begin{figure}
	\includegraphics[width=\textwidth]{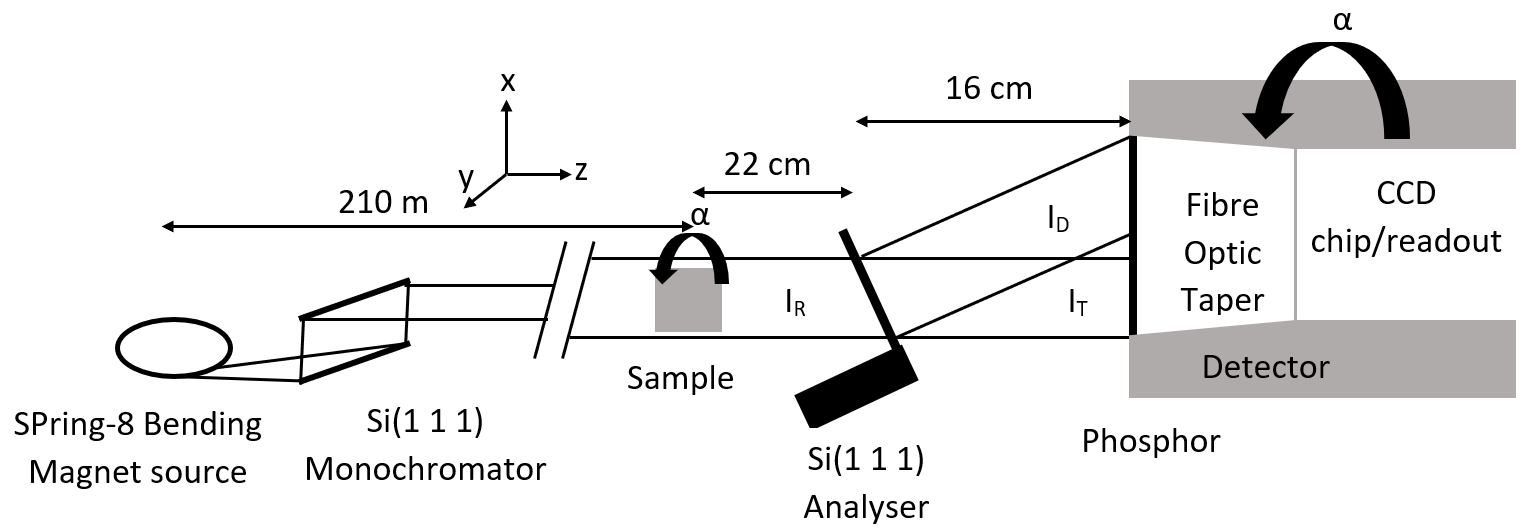}
	\caption{Inclined geometry Laue ABPCI experimental setup. The inclination was applied through rotation of the sample and detector about the optical axis.}
	\label{fig:realsetup}
\end{figure}
The aim of this paper was to apply the methodology of inclined geometry, proposed in \citeasnoun{ruthihauser2011}, to ABPCI to reconstruct 2-D phase maps and improve the 3-D reconstructions of an object’s complex refractive index using ABPCI. This 2-D phase reconstruction for ABPCI was achieved by rotating the detector and sample by an identical angle about the optical axis, while using the Laue geometry of ABPCI, as seen in Figure \ref{fig:realsetup}. 2-D ABPCI phase reconstruction has been previously achieved by \citeasnoun{modregger2007} using two analyser crystals in perpendicular directions to achieve 2-D phase sensitivity and by \citeasnoun{pavlov2004}, \citeasnoun{pavlov2005}, \citeasnoun{coan2005phase} using a variant of combined ABPCI and Propagation-Based Phase-Contrast Imaging (PBPCI). Our 2-D phase sensitive ABPCI is more straightforward and robust than the aforementioned methods \cite{pavlov2004,pavlov2005,coan2005phase,modregger2007}. For instance, we use a single crystal in a simple setup, which does not suffer from the intensity loss due to the interaction of the wavefield with an additional crystal.

\section{Theory and Methods}\label{sec:theo}

This section describes the theory and the methods using an inclined geometry Laue ABPCI setup to reconstruct the real and imaginary parts of the refractive index.

\subsection{Approximations applied to Phase-Contrast Imaging}\label{sec:applied}
The phase retrieval procedure outlined in Section (\ref{sec:phase}) is based upon the Geometrical Optics Approximation (GOA). The GOA incorporates the paraxial and projection approximations, allowing the simplification of the phase retrieval procedure by assuming smallness of the second derivative of the phase \cite{indenbom1972geometric,bushuev1996dynamicalx,bushuev1998wave,pavlov2001variant,pavlov2004,paganin2006coherent,nesterets2006}. The Laue geometry of ABPCI allows phase retrieval to be performed with two images (diffracted and transmitted) of the sample to be acquired simultaneously. Applying the GOA gives us a method using the transmitted and diffracted projections acquired from the Laue geometry setup, to separate the refraction and attenuation information \cite{ingal1995xx,bushuev1996dynamicalx,kitchen2008,kitchen2010a,kitchen2011}. Some samples have unresolvable microstructure that produces Ultra Small Angle X-ray Scattering (USAXS). The USAXS can be reconstructed using multiple image radiography that requires multiple sets of data to be recorded upon rotation of the analyser crystal \cite{oltulu2003,wernick2003,pagot2003,nesterets2006}. The multiple-image method allows the effects of refraction and USAXS to be separated and can be applied with the Laue geometry using data sets of either the transmitted or diffracted projections \cite{kitchen2010a}.
We focused on a sample that does not have any appreciable microstructure within the sample and hence produces minimal USAXS as shown in \citeasnoun{kitchen2010a}. Therefore the simultaneous dual-image Laue geometry method, neglecting USAXS, is suitable for imaging this sample. 
The size of the Borrmann triangle base (see e.g. \citeasnoun{bushuev2005}) is about 15 microns in our experiment. Therefore we used the detector with an effective pixel size of $16.2$ microns and a spatial resolution of $\sim{3}$ pixels ($\sim{50}$ microns)(i.e., larger than the Borrmann triangle base) in our experiment. Thus the Borrmann fan could not significantly affect the resolution in our experiment.

\subsection{ABPCI Phase retrieval}\label{sec:phase}
\begin{figure}
	\includegraphics[width =\textwidth]{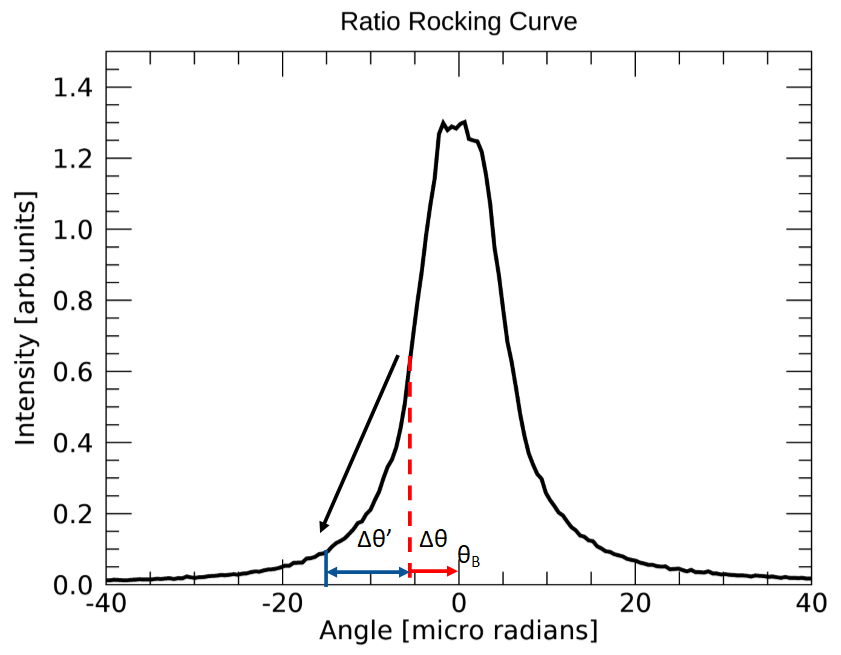}
	\vspace*{-10mm}
	\caption{Ratio RC of the diffracted RC divided by transmitted RC. The analyser crystal was positioned at the dashed red line working point on the RC through rotating it about the horizontal axis to achieve an angular shift $\Delta\theta$, shown as the red arrow from the Bragg angle $\theta_B$ position, which is placed at the origin. When an object is placed in the path of the wavefield it will cause refraction and attenuation in the wavefield propagated through the object. This changes the incident angle of the wavefield entering the analyser crystal and thus shifts it to a new position on the RC shown as the blue line with an angular shift $\Delta\theta'$ shown as the blue double arrow. We can calculate this shift from the change in intensity. These calculations will generate an intensity map and two $\Delta\theta'$ maps for every projection, as observed in Figure \ref{f9}.}
	\label{fig:RatioRC}
\end{figure}

We performed phase retrieval following a method derived by \citeasnoun{kitchen2010a} utilizing rocking curves (RCs) produced by rotating the analyser crystal. These rocking curves are produced for every pixel in the transmitted and diffracted images. We also measured the ratio of diffracted over transmitted projections without the object present in the wavefield. 
RCs can be modeled using a Taylor series, Gaussian distribution or a PearsonVII function \cite{pearson1916ix}, allowing phase retrieval to be performed. Gaussian functions are commonly used to fit the RCs as they are relatively easy to implement and accurately models the bell curve shape \cite{zhifeng2007extraction,hu2008comparison,diemoz2010absorption,arfelli2018}. However, Gaussian functions can fail at accurately modeling the peak and tails of the RC from the long slit geometry of ABPCI \cite{oltulu2003,nesterets2006}. The broadening of the RC tails from the long slit geometry is caused by scattering being integrated in the direction perpendicular to the diffraction plane \cite{suortti2013analyser}. PearsonVII functions have been shown to more accurately model the peaks and tails of the RCs \cite{kitchen2010a}. Using the phase retrieval method of \citeasnoun{kitchen2010a}, with the GOA, the transmitted ($I_T$) and diffracted ($I_D$) intensities, produced from the Laue geometry ABPCI setup, can be approximated as
\begin{equation}\label{rc1}
    I_T = I_R{T(\Delta\theta+\Delta\theta{'})},
\end{equation}
and
\begin{equation}\label{rc2}
    I_D = I_R{D(\Delta\theta+\Delta\theta{'})},
\end{equation}
respectively. Here $I_R$ is the intensity of the refracted beam incident on the crystal, $T(\Delta\theta+\Delta\theta{'})$ and $D(\Delta\theta+\Delta\theta{'})$ are the angularly dependent diffraction and transmission coefficients, $\Delta\theta$ is the deviation from the Bragg angle and $\Delta\theta{'}$ is the shift caused by refraction in the object as seen in Figure \ref{fig:RatioRC}.
We can obtain an expression independent of $I_R$ by dividing Eqn (\ref{rc2}) by Eqn (\ref{rc1}) to obtain
\begin{equation}\label{rc3}
    \frac{I_D}{I_T} = \frac{D(\Delta\theta+\Delta\theta{'})}{T(\Delta\theta+\Delta\theta{'})}.
\end{equation}
This ratio RC is used to perform phase retrieval. We modelled the RCs with a PearsonVII function given by \citeasnoun{hall1977} of the form
\begin{equation}\label{eqn70000}
y=c\big[1+(x-\tilde{x})^2/(ma^2)\big]^{-m}.
\end{equation}
Here $c$ defines the amplitude, $x$ is the independent variable, $\tilde{x}$ is the centroid, $m$ is the rate of decay of the tail and $a$ and $m$ determine the profile of the curve. This function can be adapted to the type of bell curve by modifying the $m$ such as the Lorentzian ($m=1$), the modified Lorentzian ($m=2$) and Gaussian ($m\rightarrow\infty$). We can apply this model to the ratio RC to give
\begin{equation}\label{rc6}
    \frac{I_D}{I_T} = c[1+(\Delta\theta+\Delta\theta{'})^2/(ma^2)]^{-m}.
\end{equation}
 We can rearrange Eqn (\ref{rc6}) for $\Delta\theta+\Delta\theta{'}$ to give
\begin{equation}\label{eq:1333}
\Delta\theta+\Delta\theta{'} = \pm{a}\sqrt{m[(cI_T/I_D)^{1/m}-1]},
\end{equation}
which is an expression for the angular deviation with respect to the Bragg angle position of the wavefield incident upon the analyser crystal.
Furthermore, we can rearrange Eqns (\ref{rc1}) and (\ref{rc2}) for $I_R$ to give
\begin{equation}\label{rc4}
    I_R = \frac{I_T}{{T(\Delta\theta+\Delta\theta{'})}},
\end{equation}
\begin{equation}\label{rc5}
    I_R = \frac{I_D}{{D(\Delta\theta+\Delta\theta{'})}}.
\end{equation}
This gives us potentially two relations to calculate the intensity contrast of the x-ray wavefield. We can fit an inverted PearsonVII function to the transmitted RC such that
\begin{equation}\label{eq:13333}
I_T=I_RT(\theta)=I_R\{f-d[1+\theta^2/(nb^2)]^{-n}\}.
\end{equation}
The PearsonVII coefficients $b$, $d$, and $n$ are equivalent variables to $a$, $c$ and $m$ in Eqn (\ref{rc6}) and applied to avoid confusion between the two fitted RCs with $f$ being the only unique coefficient.
\subsection{Phase Retrieval using an Inclined Geometry}\label{sec:app}
\begin{figure}
	\includegraphics[width=\textwidth]{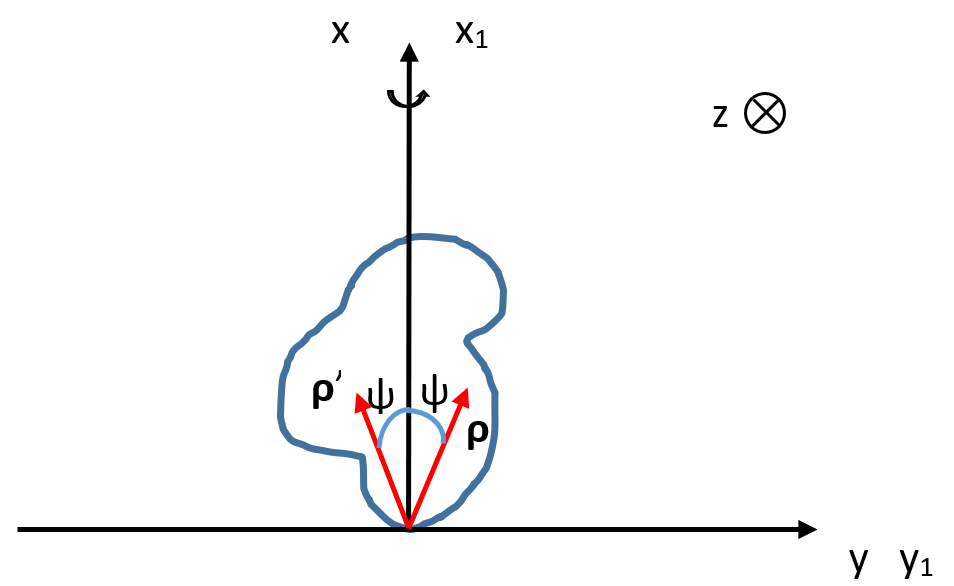}
	\vspace*{-10mm}
	\caption{Non-inclined geometry of phase gradient, $\bm{\rho}$, with its corresponding $180^\circ$ projection, $\bm{\rho}'$, where the two coordinate systems $x$, $y$ and $x_1$, $y_1$ are the vertical and horizontal vectors for the object and detector, and analyser crystal, respectively, are equivalent and $z$ is the propagation direction of the x-ray wavefield going into the page. In this setup the analyser crystal is only sensitive to variations of the phase in the $x_1$ direction.}
	\label{fig:noninc}
\end{figure}

\begin{figure}
	\includegraphics[width=\textwidth]{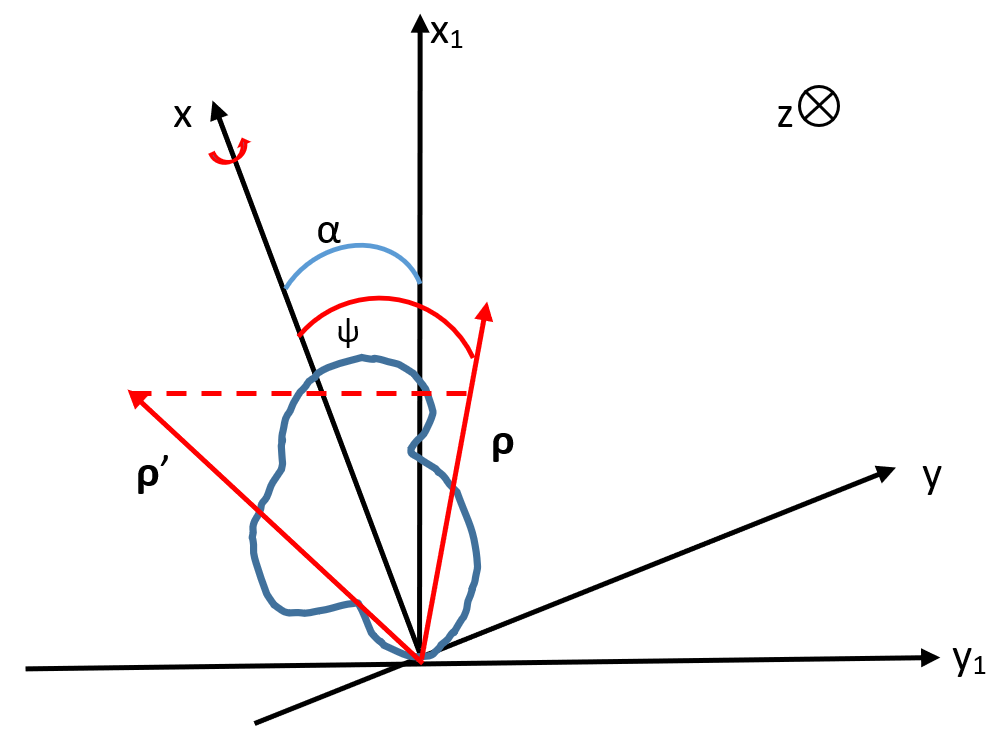}
	\vspace*{-10mm}
	\caption{Inclined geometry where the object, detector and, therefore, $(x,y)$ coordinate system has been rotated by $\alpha$ anticlockwise with respect to the $(x_1,y_1)$ coordinate system about the optical axis $z$. While the analyser crystal is still only sensitive to the phase variations in the $x_1$ direction in the $(x_1,y_1)$ coordinate system, it is sensitive to both the $x$ and $y$ components of the gradient of phase. This allows a 2-D phase gradient to be reconstructed from comparison of the two projections $\bm{\rho}$ and $\bm{\rho}'$ as they provide unique information in the inclined geometry setup.}
	\label{fig:inc}
\end{figure}

The phase shift of the wave, propagated through the sample, with spatial coordinates defined in Figure \ref{fig:noninc}, can be expressed in the form of \cite{paganin2006coherent}
\begin{equation}\label{eq:14}
\Phi=-\int{k}\delta(x,y,z)dz.
\end{equation}
Here $\delta$ is defined as the refractive index decrement of a sample and $k=2\pi/\lambda$ is the wavenumber. Furthermore, $\delta$ is related to the absorptive properties of the sample, $\beta$, and the refractive index, $n$, through \cite{james1954}
\begin{equation}\label{inteqre}
n=1-\delta+i\beta.
\end{equation}
We can measure the appropriate components of the phase gradient
\begin{equation}\label{eq:15}
\frac{\partial \Phi}{\partial x} = -k\Bigg[\frac{\partial }{\partial x}\int\delta(x,y,z)dz\Bigg],
\end{equation}
by looking at the angular shift in the rocking curve, $\Delta\theta{'}$, caused by the object in the beam
\begin{equation}\label{eq:16}
\Delta\theta{'} =-\frac{1}{k}\frac{\partial \Phi}{\partial x}. 
\end{equation}
It should be noted that Eqns (\ref{eq:15}), (\ref{eq:16}) and Figure \ref{fig:noninc} illustrate the situation when both the $x$ and $x_1$ axes are parallel to the direction of the 1-D sensitivity of the analyser crystal. We applied an $\alpha=8^\circ$ inclination of the object and detector clockwise following the x-ray propagation direction and from this a two dimensional phase gradient can be reconstructed. This is done through differential phase images from opposing projections being combined that will produce both components of the phase gradient vector $\frac{\partial \Phi}{\partial x}$ and $\frac{\partial \Phi}{\partial y}$ by retrieving $\Delta\Tilde{\theta{'}}$ and $\Delta\hat{\theta}{'}$. $\Delta\hat{\theta}{'}$ is the rocking curve shift at the $\tilde{\phi} + 180^{\circ}$ projection, while $\Delta\Tilde{\theta{'}}$ is the angular shift at the $\tilde{\phi}$ projection. In our chosen geometry $\Delta\hat{\theta{'}}$ corresponds to the projection of $\bm{\rho{'}}$ on the $x_1$ axis (see Figure \ref{fig:inc}) and $\Delta\Tilde{\theta{'}}$ corresponds to the projection of $\bm{\rho}$ on the $x_1$ axis. The equations for these two angular shifts will be of the form
\begin{equation}\label{eq:17}
\Delta\Tilde{\theta{'}}=-\frac{\kappa_1\bm{\rho}_{x}+\kappa_2\bm{\rho}_{y}}{k},
\end{equation} 
\begin{equation}\label{eq:18}
\Delta\hat{\theta}{'}=-\frac{\kappa_3\bm{\rho}_{x}+\kappa_4\bm{\rho}_{y}}{k},
\end{equation} 
where $\kappa_1$, $\kappa_2$, $\kappa_3$ and $\kappa_4$ are constants accounting for the rotation of the detector and sample at both projections and $\bm{\rho}_{x}$, $\bm{\rho}_{y}$ are the components of the phase gradient in the x and y direction, respectively. To calculate these values we need to consider the effect of $\alpha$ inclination to the object and detector by looking at both the non-inclined and inclined geometries.

For clarity, rather than using the angle $\Delta\theta{'}$, let us consider that we have a phase gradient, $\bm{\rho}$, and its respective $180^\circ$ projection, $\bm{\rho}'$, for the non-inclined geometry (see Figure \ref{fig:noninc}). We can derive a simple expression for the $x$, $x_1$ and $y$, $y_1$ components of $\bm{\rho}$ using simple trigonometry, see Figure \ref{fig:noninc}
\begin{equation}\label{eq:19}
\bm{\rho}_{x_1}=|\bm{\rho}|\cos(\psi)=\bm{\rho}_{x},
\end{equation}
\begin{equation}\label{eq:20}
|\bm{\rho}_{y_1}|=|\bm{\rho}|\sin(\psi)=|\bm{\rho}_{y}|.
\end{equation}
Here $|\bm{\rho}|$ is the magnitude of phase gradient, $x$ and $y$ are the axes for the object and detector, $x_1$ and $y_1$ are the axes for the analyser crystal, which is only sensitive to the phase variations in the $x_1$ direction and $\psi$ is the angle between the vector $\bm{\rho}$ and the $x$ axis.

The two coordinate systems are equivalent, as shown in Figure \ref{fig:noninc}. However, if we rotate the detector and sample by an angle $\alpha$ anticlockwise along the path of the wavefield, the $x$, $y$ coordinates and the orientation of the object will change with respect to coordinates $x_1$, $y_1$, as seen in Figure \ref{fig:inc}. We can again derive expressions for components of $\bm{\rho}$ and utilize the cosine law $\cos(A+B) =\cos{A}\cos{B}-\sin{A}\sin{B}$ to give
\begin{equation}\label{eq:21}
\begin{split}
\bm{\rho}_{x_1} =|\bm{\rho}|\cos(\psi-\alpha)=|\bm{\rho}|[\cos\psi\cos(\alpha)+\sin{\psi}\sin(\alpha)]=-k\Delta\Tilde{\theta{'}},
\end{split}
\end{equation}
\begin{equation}\label{eq:22}
\bm{\rho}'_{x_1}=|\bm{\rho}|\cos(\psi+\alpha)=|\bm{\rho}|[\cos\psi\cos(\alpha)-\sin\psi\sin(\alpha)]=-k\Delta\hat{\theta}{'},
\end{equation}
from Eqn (\ref{eq:16}). From here we can add Eqns (\ref{eq:21}) and (\ref{eq:22}) to obtain
\begin{equation}\label{eq:23}
-k(\Delta\Tilde{\theta{'}}+\Delta\hat{\theta}{'})=2\bm{\rho}_{x}\cos(\alpha).
\end{equation}
Here we used $\bm{\rho}_{x}$, from following Figure \ref{fig:inc}, as
\begin{equation}\label{eq:41}
\bm{\rho}_{x}=|\bm{\rho}|\cos(\psi),
\end{equation}
then rearranging Eqn (\ref{eq:23}) to obtain
\begin{equation}\label{eq:24}
\bm{\rho}_{x}=-\frac{k(\Delta\Tilde{\theta{'}}+\Delta\hat{\theta}{'})}{2\cos(\alpha)}.
\end{equation}
Similarly for subtracting Eqns (\ref{eq:21}) and (\ref{eq:22}) we get
\begin{equation}\label{eq:25}
-k(\Delta\Tilde{\theta{'}}-\Delta\hat{\theta}{'}) =2\bm{\rho}_{y}\sin(\alpha).
\end{equation}
Here we used $\bm{\rho}_{y}$, from following Figure \ref{fig:inc}, as
\begin{equation}\label{eq:42}
\bm{\rho}_{y}=|\bm{\rho}|\sin(\psi),
\end{equation}
then after rearranging Eqn (\ref{eq:25}) we obtain
\begin{equation}\label{eq:26}
\bm{\rho}_{y}=-\frac{k(\Delta\Tilde{\theta{'}}-\Delta\hat{\theta}{'})}{2\sin(\alpha)}.
\end{equation}
Therefore, going back to Eqns (\ref{eq:17}) and (\ref{eq:18}) the expression for the coefficients is given by
\begin{equation}\label{eq:27}
\kappa_1=\cos(\alpha),
\end{equation}
\begin{equation}\label{eq:28}
\kappa_2=\sin(\alpha),
\end{equation}
\begin{equation}\label{eq:29}
\kappa_3=\cos(\alpha),
\end{equation}
\begin{equation}\label{eq:30}
\kappa_4=-\sin(\alpha).
\end{equation}
This method will allow the reconstruction of a 2-D phase gradient with the additional phase information gathered using an inclined geometry. This is achieved by mirroring $\Delta\hat{\theta}'$ about the vertical axis so that it matches with its opposing plane. These planes will provide different information about the object that can be extracted and used in tomographic reconstruction.

\section{Experimental Setup}\label{sec:exp}

This experiment was performed in hutch 3 of beamline 20B2 in the Medium-length Beamline Facility at the SPring-8 radiation facility (Japan) using a mounted perspex phantom as a sample. The imaged cylindrical perspex phantom was $12.75$\,mm in diameter with four $1.02$\,mm diameter cylindrical holes in the top of the phantom. Two of these holes were filled with aluminium and teflon pins with $1.02$\,mm diameter each with a cap on the top, while the other two were left empty. This phantom was discussed in greater detail in \citeasnoun{Beltran2010}. We employed an inclined Laue geometry ABPCI experimental setup as observed in Figure \ref{fig:realsetup}.

Following from left to right in Figure \ref{fig:realsetup} we have the synchrotron set to produce x-ray wavefields approximately 210\,m away from the sample. The x-ray wavefields then interacted with a double-bounce monochromator in a non-dispersive setup. This consisted of two parallel Si$(1 1 1)$ crystals that monochromatize the x-rays yielding a $26$\,keV monochromatic wavefield with energy bandwidth $\Delta{E}/E\approx{10^{-4}}$\, \cite{goto2001construction}. This x-ray wavefield then interacted with the object, with intensity $I_R$ just after the object that was rotated $\alpha=8^\circ$ clockwise about the optical axis following the propagation direction of the x-ray wavefield. \citeasnoun{ruthihauser2011} applied an ideal $45^\circ$ rotation of the two gratings in their experimental setup, while the tilt stages available for use in our experiment were limited to $8^\circ$. Because of the small $\alpha=8^\circ$ inclination angle we applied, our setup will still be predominantly sensitive to phase effects in the $x$ direction. The x-ray wavefield then traveled $22$ cm from the sample before being incident on the near perfect Si$(1 1 1)$ analyser crystal in the Laue geometry. 

This analyser crystal consisted of a nominally $100$\,$\mu$m thick silicon wafer that was connected at the base to a monolithic silicon slab. The interaction between the x-ray wavefield and the analyser crystal caused the x-ray wavefield to be simultaneously diffracted and transmitted with respective intensities $I_D$ and $I_T$. The diffracted x-ray wavefield then propagated with an angle $2\theta_B=8.722^\circ$ with respect to the propagation direction of the incident wavefield \cite{stepanov2004x,stepxray}.

The data from these separated beams were then gathered by a $4000\times{2672}$ pixel Hamamatsu CCD camera (C9399-124F), with a tapered fibre optic bonded to the CCD chip and the $20$\,$\mu$m thick gadolinium oxysulfide ($Gd_2O_2S:Tb^+;P43$) phosphor. The CCD detector with native pixel size of $9$\,$\mu$m was converted to an effective pixel size of $16.2$ $\mu$m by the 1.8:1 taper ratio. The CCD detector was positioned $16$ cm away from analyser crystal and was also rotated $8^\circ$ clockwise following the propagation direction of the x-ray wavefield.
\begin{figure}
	\includegraphics[width=0.8\textwidth]{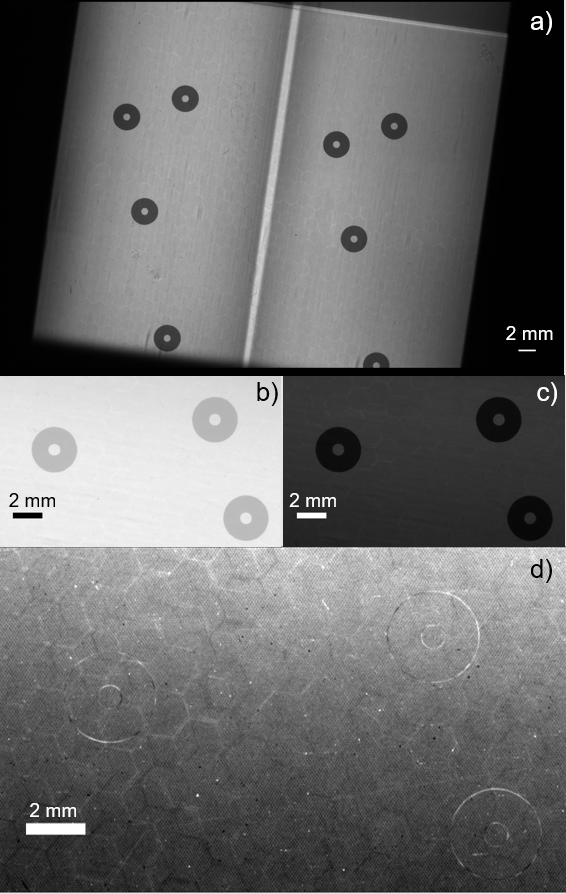}
	\caption{a) Transmitted (left) and diffracted (right) projections of three gold foil fiducial markers captured by a single exposure. Separated and aligned b) diffracted and c) transmitted images produced using the positions of the three fiducial markers in the projections. We can check the quality of the alignment by producing an image of the d) difference between diffracted over transmitted projections. Any misalignment between these two images will be visible through bright and dark arcs around the fiducial markers, making them stand out from the background. Only slight imperfections in the alignment can be seen. The `chicken wire' structure is produced from the fibre optic taper in the detector the wavefield travels through before incident upon the CCD chip.}
	\label{fig:first}
\end{figure}
\begin{figure}
	\includegraphics[width=1\textwidth]{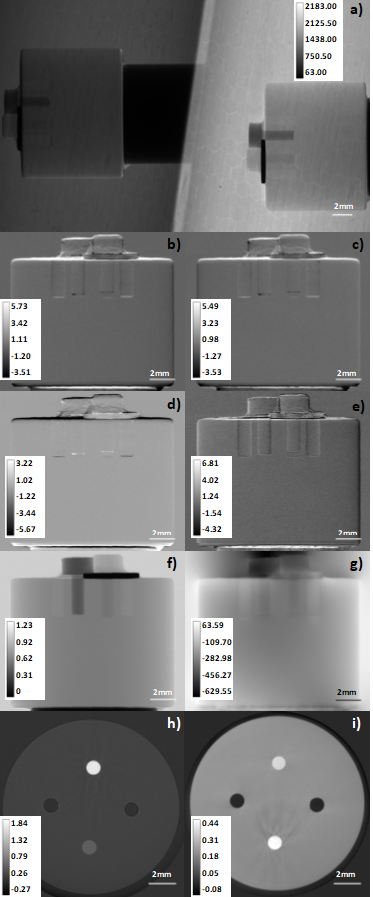}
    \label{f9}
    \vspace*{-10mm}
    \caption{Caption on next page}
\end{figure}
\addtocounter{figure}{-1}
\begin{figure}
    \vspace*{-10mm}
	\caption{Maps generated throughout the experimental procedure beginning with the a) raw tomographic data, projection number 362, showing the diffracted (left) and transmitted (right) intensities in this single projection that needs to be separated and aligned. We then perform phase retrieval to obtain maps b) $\Delta\Tilde{\theta{'}}$ of the change of the angle of incidence upon the analyser crystal and c) $\Delta\hat{\theta}{'}$, the $180^\circ$ equivalent of $\Delta\Tilde{\theta{'}}$ in microradians. From this we split $\Delta\Tilde{\theta{'}}$ and $\Delta\hat{\theta}{'}$ into the d) vertical and e) horizontal components of the phase gradients, divided by the wavenumber, in microradians. From the transmitted projection we obtain a map of f) intensity, while performing 2-D integration using d) and e) to calculate the g) phase map in radians. We then performed tomographic reconstruction using the intensity and phase maps to calculate the h) $\beta$ maps ($\times{10^{-9}}$) , and i) $\delta$ maps ($\times{10^{-6}}$), respectively.}
\end{figure}
\subsection{Diffracted and Transmitted Image Alignment}\label{sec:align}
The data was dewarped using triangular interpolation to correct for the distortion caused by the fibre optic taper \cite{kitchen2010a,islam2010}. We applied a Laue geometry ABPCI method that allows one the simultaneous acquisition of diffracted and transmitted images of the object captured by a single CCD detector similar to \cite{kitchen2011}, see Figure \ref{fig:realsetup}.
The alignment of the transmitted and diffracted images was achieved using three gold foil disks placed in the object plane, as seen in Figure \ref{fig:first}. Upon locating the central coordinates of the foils, we used the three pairs to align the images via the affine transformation described by \citeasnoun{kitchen2011}. From Figure \ref{fig:first}, we can see the alignment procedure appears to fairly successfully align the transmitted and diffracted projections as the aligned and subtracted gold foils markers blend in well with the background, as seen in Figure \ref{fig:first}d).

\section{Results}

Following the phase retrieval procedure discussed in Section (\ref{sec:theo}), maps of the object were obtained, as shown in Figure \ref{f9}, with the analyser crystal positioned at a working point of $50\%$ peak intensity on the left side of the RC. Beginning with the raw data a) we have the transmitted and diffracted phase contrast images on the right and left hand sides of the image, which must be separated and aligned, as discussed in Section (\ref{sec:align}). We then fit rocking curves with a PearsonVII function to the ratio and diffracted projections with no object present in the beam for each pixel in the images. We used these fitted rocking curves with the transmitted and diffracted projections to calculate the b) $\Delta\Tilde{\theta{'}}$, c) $\Delta\hat{\theta}{'}$ and f) the attenuation contrast image.

We then split $\Delta\Tilde{\theta{'}}$ and $\Delta\hat{\theta}{'}$ into the d) vertical and e) horizontal components of the phase gradients, which were then integrated to calculate the phase map g). We then performed $180^\circ$ CT filtered back projection reconstruction using the attenuation contrast and corrected phase maps (see section 4.1) to produce 3-D reconstructions of $\beta$ and $\delta$, respectively. Reconstructions were obtained for a slice of the reconstructed $\delta$ and $\beta$ maps in Figure \ref{f9} i) and h) as shown in Table \ref{table1}. The uncertainties were calculated by taking the standard deviation over some area around the reference point in the slice \cite{rasband1997imagej}.

The measured $\beta$ values are in a good agreement with the theoretical ones. However, the $\delta$ values are all approximately a factor of two smaller than the theoretical ones.
\begin{table}
	\label{table1}
	\caption{Reconstructed $\delta$ and $\beta$ values for media present in the reconstructed phantom with theoretical values obtained from \citeasnoun{henke1993x}. Note: Al = Aluminium, PMMA = Perspex, The = Theoretical, Mea = Measured.}
	\begin{tabular}{lllll} 
		&$\beta_{The}$    & $\beta_{Mea}$        & $\delta_{The}$       & $\delta_{Mea}$\\
		\hline
		Al   & $1.5\cdot{10^{-9}}$ & $1.6\pm0.1\cdot{10^{-9}}$ & $8.0\cdot{10^{-7}}$ & $4.3\pm0.1\cdot{10^{-7}}$\\
		PMMA     & $1.4\cdot{10^{-10}}$ & $1.5\pm0.3\cdot{10^{-10}}$ & $3.9\cdot{10^{-7}}$ & $2.1\pm0.1\cdot{10^{-7}}$\\
		Teflon      & $3.7\cdot{10^{-10}}$ & $4.0\pm0.3\cdot{10^{-10}}$ & $6.5\cdot{10^{-7}}$ & $3.6\pm0.1\cdot{10^{-7}}$\\ 
	\end{tabular}
\end{table}

\begin{figure}
	\includegraphics[width=1\textwidth]{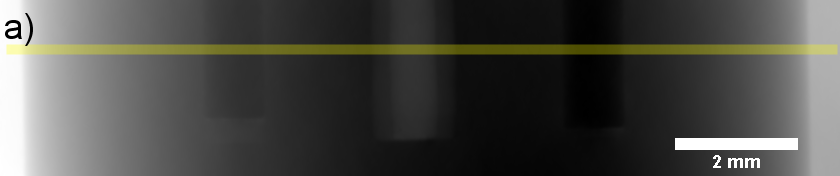}
	\includegraphics[width=1\textwidth]{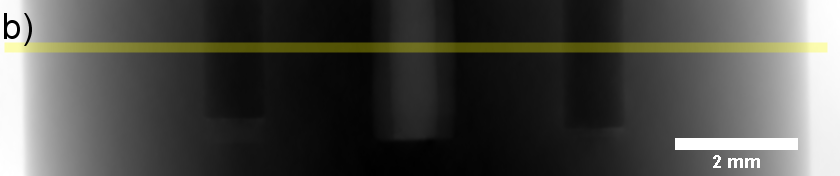}
	\includegraphics[width=1\textwidth]{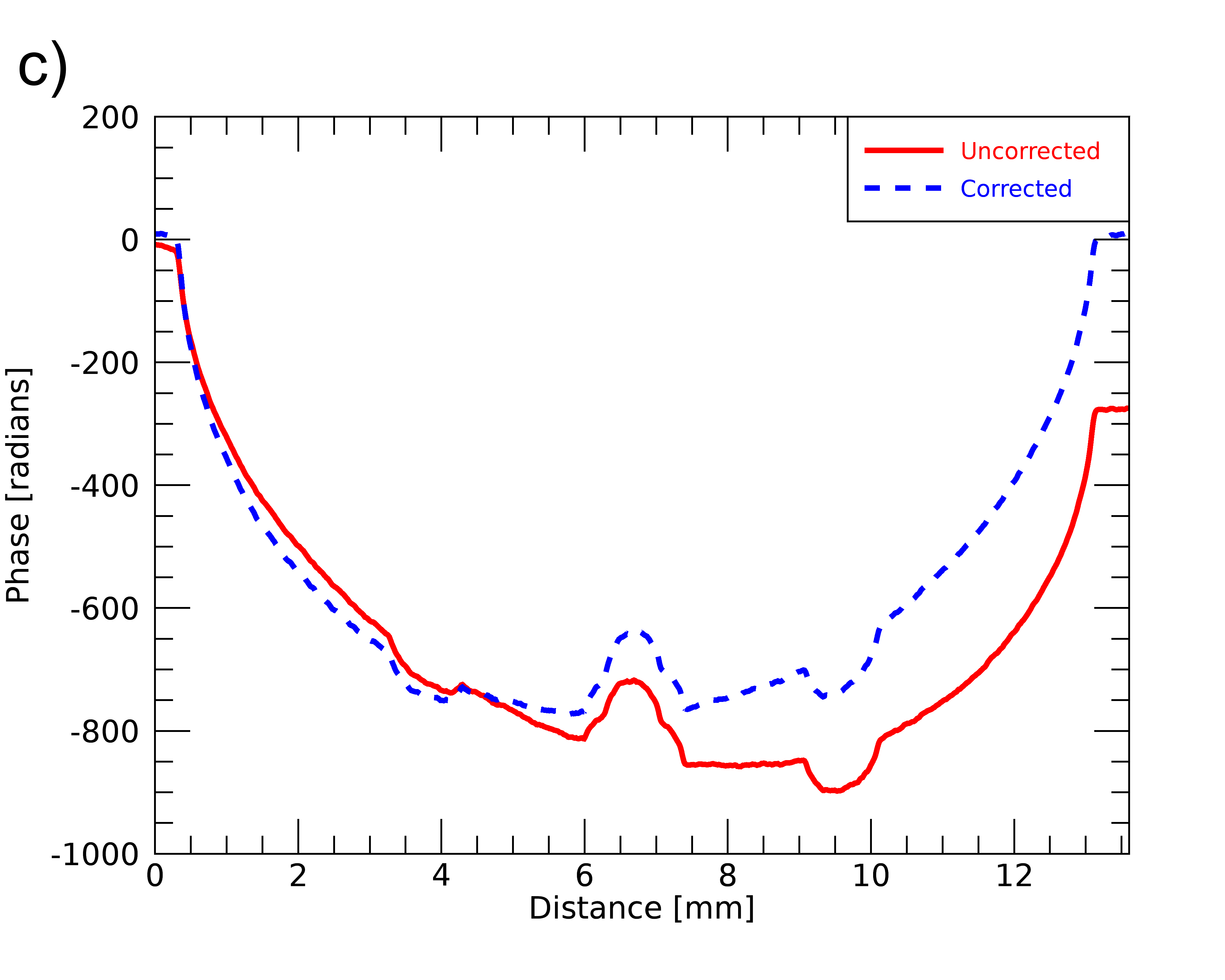}
	\vspace*{-10mm}
	\caption{Phase maps a) without and b) with the linear trend correction applied through measuring the linear gradient between the left and right edge of the phantom then dividing it through the entire phase map. c) Plots for the uncorrected and corrected phase maps given by the red solid line and blue dashed line, respectively, approximately show this linear trend.}
	\label{fig:linear}
\end{figure}

\subsection{Corrections}\label{sec:corr}

 Under plane wave illumination the phase outside the object should be approximately constant. However, Figure \ref{fig:linear} shows large low frequency phase gradients are present across the images. This comes from the 2-D integration process to calculate the phase map, which amplifies low frequency noise in the image. A linear ramp correction was applied to the phase map as the phase of one side of the phantom was underestimated with respect to the other. This linear correction was applied during the phase retrieval process to all the phase maps in order to make the sides of the phantom have the same phase value. Figure \ref{fig:linear}a) and b) show phase maps and their plots in c) with and without the correction applied. We see that for the uncorrected phase map the right hand side is lower than the left hand side and we also see a slope in the parabolic shape. This is corrected by estimating the linear ramp through measuring the phase values on the left and right hand side of the phantom for each line, normalising, then dividing the phase map by the linear ramp. The linear ramp of the phase behaves inhomogeneously, changing in both magnitude and sides of the phantom over the sequence of acquired phase maps. Causes of these approximately linear trends are discussed in Section 5.

\section{Discussion}

Qualitatively, our reconstructions of the $\beta$ and $\delta$ distributions shown in Figure \ref{f9}h) and i) are in excellent agreement with expectation and there are minimal artefacts seen in the reconstructions, despite having to correct the phase maps. The application of the correction procedure to the phase retrieval process has provided high contrast and high resolution reconstructions of the object even though it is not the most effective CT filtering method. The quantitative measures of the attenuation properties are in excellent agreement with theoretical predictions, as shown in Table \ref{table1}. The underestimation of the $\delta$ values, however, is most likely due to the underestimation of the phase gradient. Inaccuracies in the phase gradient maps lead to low frequency artefacts of the phase maps. This was partially corrected by applying the linear correction. These inaccuracies can arise from (1) the failure of the GOA at boundaries due to the high phase gradient, (2) imperfect alignment of the transmitted and diffracted projections and (3) the shallow $8^\circ$ inclination applied.

We explore these issues, beginning with point (1). The GOA assumes slow variations of the phase as the wavefield propagates through the sample. This assumption may break down at the boundaries between PMMA and air, where the refractive index difference is quite large. This could be fixed by submerging the sample in a fluid with similar refractive properties to PMMA such as paraffin, which would reduce the change in the phase gradient. \citeasnoun{ruthihauser2011} obtained results for rat cerebellum submerged in paraffin, using the inclined geometry, of $\delta=4\times{10^{-7}}$ with small discrepancies when compared to the theoretical value of $3.52\times{10^{-7}}$ from \cite{Ts,brennan1992suite,white1988tissue,chantler2003x,zschornack2007handbook,stepanov2004x,stepxray,stevenson1993x}. Whereas, \cite{ruthihauser2011,rutishauser2013} demonstrated a large discrepancy of about one order of magnitude  between the reconstructed and theoretical values while imaging a cylindrical PMMA phantom in air with a photon energy of 25\,keV $(\Delta{E}/E)\approx{2\%}$. The $\delta$ value for perspex was reconstructed to be $0.4\times{10^{-7}}$ compared to the theoretical value of $3.9\times{10^{-7}}$ \cite{henke1993x}. However, \citeasnoun{kitchen2010a} showed good agreement between the theoretical and reconstructed values of the function $\delta$ obtained for a PMMA block in air with cylindrical cavities, with a photon energy of $26$\,keV using a 1-D phase sensitive Laue ABPCI setup. In that study, the PMMA block was positioned in such a manner that the direction of the phase gradient produced by the cylinder was aligned with the direction of maximum sensitivity of the analyser crystal used. Therefore, part of our deviation from the theoretical value may result from the restricted angle ($8^\circ$) by which we could rotate the sample and detector.

Following with point (2), any misalignment between the transmitted and diffracted projections can result in significant inaccuracies in the reconstructed phase gradients. However, Figure \ref{fig:first} shows that projections appear to be relatively well aligned. It is possible that our alignment method needs further improvements in order to more accurately reconstruct the 2-D phase gradient map as even subpixel misalignments can have a significant effect \cite{kitchen2011}. It is important to note that the $\beta$ values were calculated from a single set of $180^\circ$ projections, while the $\delta$ values used two sets of projections, the second coming from mirroring the $\Delta\hat{\theta}{'}$  projections, as described previously. This could suggest that complications may have occurred when utilizing the mirrored projections causing the observed discrepancy.

Finally, following point (3), we recall that the analyser crystal is only sensitive to the component of the phase gradients lying in the plane of diffraction of the analyser crystal. For our experimental setup, this was in the vertical direction. Our mechanical restriction to $8^\circ$ inclination did allow some information from the horizontal direction to be obtained, but with low amplitude and relatively high noise compared to the vertical component, as seen by comparing Figure \ref{f9} d) and e). This horizontal component information is then amplified through division of $2\sin(8^\circ)=0.28$, as shown in Eqn (\ref{eq:26}), while the vertical component information is decreased through division of $2\cos(8^\circ)=1.98$, as shown in Eqn (\ref{eq:24}). Additional experiments need to be undertaken to test these speculations to determine the primary source of errors, beginning with increasing the inclination of the applied inclined geometry. The fact that our results have the correct order of magnitude for the $\delta$ value is therefore encouraging given the small inclination angle of just 8 degrees. We anticipate that a larger inclination angle of $30^\circ-60^\circ$ will lead to reconstructions with values more closely matching the theoretical values. This larger inclination will allow more information from the horizontal axis to be acquired and used in the integral to calculate the phase.

\section{Conclusion}

We applied an inclined geometry method in order to achieve 2-D phase sensitivity for ABPCI in a Laue geometry setup through rotation of the object and detector by $8^\circ$ clockwise folowing the x-ray wavefield propagation direction. Our measured $\beta$ values were in excellent agreement with the theoretical ones. The measured $\delta$ values were qualitatively correct and had the correct order of magnitude, but the measured values were approximately a factor of two less than the theoretical values. Considering the small angle inclination of the crystal relative to the sample stage and detector ($8^\circ$ compared to the ideal $45^\circ$ degrees), the results are encouraging. The discrepancy between the measured and theoretical $\delta$ values could also be due to GOA condition breaking down or slight misalignment of the transmitted and diffracted images with additional experiments required in order to confirm the source of error and obtain more accurate results. 

\ack{\textbf{Acknowledgements}}

The synchrotron radiation experiments were performed at Beamline BL20B2 of SPring-8 with the approval of the Japan Synchrotron Radiation Research Institute (JASRI) (Proposal 2012B1315). We acknowledge travel funding provided by the International Synchrotron Access Program (ISAP) managed by the Australian Synchrotron, part of ANSTO (AS/IA124/6149). We acknowledge Timur Gureyev, David M. Paganin and Iain M. Young for their work on the journal article, funding, planning and discussion of the proposal for this experiment. MJK is funded by an ARC Future Fellowship (FT160100454).

\bibliographystyle{iucr}

\end{document}